\documentstyle[12pt]{article}
\def\beqa{\begin{eqnarray}}
\def\eeqa{\end{eqnarray}}
\def\beq{\begin{equation}}
\def\eeq{\end{equation}}

\def\pa{\partial}
\def\alp{{\alpha}}

\def\ie{{\it i.e. }}
\def\eg{{\it e.g. }}

\def\pr{{\it Phys. Rev.}\ }

\def\apj{{\it Astrophys. J.}\ }
\def\aj{{\it Astron. J.}\ }

\begin{document}
\def\bib#1{[{\ref{#1}}]}

\title{\normalsize\bf Higher Order Corrections in  Gravitational Microlensing}

\author{\normalsize
S. Capozziello$^{1,2}$\thanks{E-mail:capozziello@vaxsa.csied.unisa.it},
G. Lambiase $^{1,2}$\thanks{E-mail:lambiase@vaxsa.csied.unisa.it}, 
G. Papini$^{3}$\thanks{E-mail:papini@cas.uregina.ca},
G. Scarpetta$^{1,2}$\thanks{E-mail:scarpetta@vaxsa.csied.unisa.it}  \\
{\small \em $^{1}$ Dipartimento di Scienze Fisiche ``E. R. Caianiello'',} \\
{\small \em Universit\`{a} di Salerno, I-84081 Baronissi, Salerno, Italy.} \\
{\small \em $^{2}$ Istituto Nazionale di Fisica Nucleare, Sezione di Napoli, Italy} \\
{\small \em $^{3}$ Department of Physics, University of Regina, Sask.,
 S4S 0A2 Canada.}}
\date{}
\maketitle
\begin{abstract}
We show that, in cosmological microlensing,
corrections  of order
$v/c \sim \Delta\lambda/\lambda$,
to the deflection angle
of light beams from a distant source are not negligible and that all microlensing quantities  should be corrected up to this order
independently of the cosmological model used.
\end{abstract}

\vspace{5. mm}

keywords: microlensing, redshift, cosmology.

\vspace{10. mm}

\noindent In the last decade, gravitational lensing
has become  one of the most powerful tools in
astrophysics and cosmology in studies about
the presence and mass distribution of dark matter in the Universe
\cite{schneider},\cite{ehlers},\cite{peebles}.
It affords, in principle, estimates of the gravitational mass of all large scale
structures, from galaxies to super clusters, and, in the specific
application called {\it microlensing}, it
 can be used
to search for {\it Massive Astrophysical Compact
Halo Objects} (MACHOs) \cite{paczynski}.
These objects are considered to be the main constituents of the dark halo of
spiral galaxies (of our Galaxy, in particular) and,
theoretically, could range in mass from $10^{-8}\div 10^{2}M_{\odot}$.
MACHOs
could therefore represent planets, brown dwarfs, or
massive black holes \cite{carr}.
The fundamental problem is how lensing by a point--like mass can be detected.
Unless the lens is very massive ($M>10^{6}M_{\odot}$),
the angular separation of  two
images, usually produced by a point-like lens, is too small to be resolved.
The angular separations of images
are, in fact, of the order $\sim 10^{-3}\div 10^{-6}$ arcsec,
hence the term {\it microlensing}.
However, even when detecting  multiple images is impossible, the
magnification can still be seen if the lens and the source move
relative to each other. This motion gives rise to
a lensing--induced time dependence
of the source luminosity \cite{chang}.
 Since the effect
was first observed for the quasars QSO 2237+0305 and QSO 0957+561
\cite{irwin},\cite{schild}, one must now distinguish {\it galactic}
microlensing from {\it extragalactic} or {\it cosmological} microlensing.
In the first case, the light sources are stars and the angular separations
involved are $\sim 10^{-3}$arcsec. In the second case, the sources are
very distant quasars and the angular separations involved are
$\sim 10^{-6}$arcsec. In both instances the term ``microlensing" is used.
In this work we are exclusively concerned with cosmological microlensing. The principle of microlensing is quite simple.
If the closest approach between a point mass
lens and a source is equal to or less  than $\theta_{E}$, the Einstein  angular
radius, the peak magnification in the lensing--induced light curve
results in a brightness
enhancement (\eg $\sim 0.3$ magnitudes),
which can be easily
detected. As discussed below, the Einstein angular radius $\theta_{E}$
is a feature of the lens-source system and provides the natural angular
scale required to describe the lensing geometry. It gives, in fact, the typical angular separation among the single images, while for
axisymmetric lens--source--observer systems, it gives the aperture of the
circular bright image, called {\it Einstein ring}. Geometrically, the Einstein ring, can be defined in any
case, even when a luminous circular image is not produced.
However, sources which are
closer than $\theta_E$ to the optical axis experience a strong lensing effect
and are hardly magnified, while sources that are located well outside
the Einstein ring are weakly magnified. Therefore,
for a lot of lens models,
the Einstein ring represents the boundary between the zones where sources
are strongly magnified or multiply--imaged and those where they are softly
magnified or singly--imaged \cite{ehlers}.

The first detection proposal \cite{paczynski} consisted in monitoring millions of stars
in the
{\it Large Magellanic Cloud}
(LMC), or in the bulge of the Galaxy and in searching for the corresponding magnifications.
By detecting a sufficient number of events, one could then map the distribution of (dark) stellar--mass objects in
the halo  of the
Galaxy (due to the fact that LMC is near us and the halo of our galaxy
is in between) or between the Solar System and the bulge of the Galaxy.
The two alternatives require some care in the selection of distances between
source and observer. In fact, the distance between the Sun and the center
of LMC is $\sim 55$Kpc while the distance between the Sun and the Galaxy bulge is $\sim 8.5$Kpc. This difference in size yields Einstein radii for
the selected sources which could differ by about one order of magnitude.
Furthermore, the Galaxy haloes are supposed to extend for
$\sim 50$Kpc so that the zone where MACHOs can pass is very large.
Both approaches can however be used for galactic microlensing
and for $r_{E}\sim 1\div 10$AU, source--lens--observer distances $D\sim 1\div 50$Kpc
and MACHO velocities $v\sim 100\div 500$Km $s^{-1}$
microlensing effects may be observable
\cite{paczynski}.

In the proposal of Ref. \cite{paczynski}, distinguishing
the intrinsic variable stars (which are very numerous in a normal galaxy)
from the lensing--induced variables is potentially a serious problem. Fortunately, the light curves of lensed
stars have distinctive features which can be used to separate induced variability
from intrinsic variability. For instance, the light curves are symmetric in time
and chromatic effects are absent because light deflection does not depend
on the wavelength. On the contrary, intrinsic variables have asymmetric light
curves. Furthermore, magnification produces in this case strong chromatic effects.
The probability of seeing microlensing events depends on the
{\it optical depth}, which is the probability that at any
instant in time a given source be
within the angle $\theta_{E}$ of a lens. The optical depth is the integral
over the number density $n(D_{ol})$ of lenses times the area enclosed by
the Einstein ring of each lens, \ie
\beq
\label{03}
\tau=\frac{1}{\Omega}\int dV n(D_{ol})\pi\theta^{2}_{E}\,,
\eeq
where  $dV$ is the volume of an infinitesimal spherical
shell of radius $D_{ol}$ which covers a solid angle $\Omega$.
Eq.(\ref{03}) may take a very simple form if the sources are distant
and compact. Then sources and lenses have angular sizes
smaller than $\theta_{E}$.

From the cosmological point of view, microlensing plays a fundamental role
in the determination of the density parameter $\Omega_{0}$
and the cosmological constant $\Lambda$.
It has been frequently argued that a significant fraction of the dark matter in
the Universe may be in the form of compact masses which could induce
lensing phenomena. For example, let us consider
an Einstein--de Sitter Universe with a certain constant
comoving number density of point lenses of  mass $M$. Let $\Omega_{M}$ be the
density parameter. The optical depth for lensing of sources at redshift
$z_{s}$ can be shown to be
\beq
\label{04}
\tau(z_{s})=3\Omega_{M}\left[\frac{(z_{s}+2+2\sqrt{1+z_{s}})\ln
(1+z_{s})}{z_{s}}-4\right]\,,
\eeq
which is
$\tau(z_{s})\simeq \Omega_{M}z_{s}^2/4$,
for $z_{s}\ll 1$, and
$\tau(z_{s})\simeq 0.3\Omega_{M}$,
for $z_{s}=2$. Hence the number of lensing events in a given source sample  measures the cosmological density of compact objects directly. These results depend strongly on the red--shift \cite{press}. In calculating the probability
of lensing it is important to allow for various selection effects. Lenses
magnify the observed flux, and therefore sources which are intrinsically too
faint to be observed may be lifted above the detection threshold.
 At the same time, lensing increases the solid angle within
which sources are observed so that their number density in the sky is reduced
\cite{narayan}. If the faint sources are numerous, the increase in
source number due to the apparent brightening outweighs their spatial
distribution, and the observed number of sources is increased by lensing.
This magnification bias \cite{turner} can increase the probability of lensing
for bright optical quasars.
Besides, the strong dependence of the optical depth on the red--shift
suggests that the simple weak field
and slow motion approximations (normally used in lensing theory) be
applied with care. In particular, we show below that discarding terms
of the order $c^{-3}$ in the lens equation is not justified in the high
 red--shift regime.
In general, it is not always correct to consider only the term $\Phi/c^2$ in the
refractive index $n$ because the propagation of light
in a gravitational background depends on the mass distribution and, frequently, the ray approximation does not work
\cite{peters},\cite{manzano}.

In this letter we show explicitly that some effects ensue from including
higher order approximation terms in the calculation of the light deflection by a standard lens (in particular a point--like lens).

In general relativity
the weak field approximation is defined by
\beq\label{31}
g_{\mu\nu}(x)=\eta_{\mu\nu}+h_{\mu\nu}(x),
\qquad \vert h_{\mu\nu}(x)
\vert \ll 1\,{.}
\eeq

If the distances are small with respect
to the Hubble distance $c/H_0$, one can neglect the curvature and
the expansion of the Universe and
the stress--energy tensor for perfect fluid matter is given by
\beq\label{32}
T^{\mu\nu}=(p+\rho c^2)u^{\mu}u^{\nu}-pg^{\mu\nu}\,{,}
\eeq
which,
in the approximation $\vert\vec{v}\vert \ll c$ and $p\ll \rho c^2$,
reduces to the components
\beq
\label{33}
T^{00}\simeq  \rho c^2\,,\;\;\;\;
T^{0j}\simeq  \rho c v^j \,,\;\;\;
T^{ij}\simeq \rho v^iv^j\,.
\eeq
By using the Einstein equations, one finds
\beq
\label{36}
\nabla^2h_{00}= \frac{8\pi G}{c^2}\rho\,,\,\,\,\,
\nabla^2h_{ij}= \frac{8\pi G}{c^2}\delta_{ij}\rho \,,\,\,\,\,
\nabla^2h_{0j}= -\frac{16\pi G}{c^2}\delta_{jl}\rho v^l\,,\,\,\,\,
\eeq
where $\nabla^2$ is the usual Laplacian operator.
In arriving at Eqs.(\ref{36}),
use has been made of the harmonic condition
\beq\label{39}
g^{\mu\nu}\Gamma^{\alpha}_{\mu\nu}=0\,{.}
\eeq
The integration of Eqs.(\ref{36}) yields
\beq\label{40}
h_{00}=  -\frac{2G}{c^2}\int \frac{\rho}{\vert \vec{x} -
\vec{x}^{\prime}\vert  }d^3x^{\prime} \,,\;\;
h_{ij}= -\frac{2G}{c^2}\delta_{ij}\int \frac{\rho}{\vert \vec{x} -
\vec{x}^{\prime}\vert     }d^3x^{\prime} \,,\;\;
h_{0j}= \frac{4G}{c^3}\delta_{jl}\int \frac{\rho v^l}{\vert \vec{x} -
\vec{x}^{\prime}\vert    }d^3x^{\prime} \,.
\eeq
Up to leading order in $v/c$, the metric is determined
by the gravitational potential
\beq\label{43}
\Phi(x) = -G\int \frac{\rho}{\vert \vec{x} -
\vec{x}^{\prime}\vert }d^3x^{\prime}\,
\eeq
and by a potential $V^l$
\beq\label{44}
V^l=-G\int \frac{\rho v^l}{\vert \vec{x} -
\vec{x}^{\prime}\vert }d^3x^{\prime}\,{.}
\eeq
From Eqs. (\ref{31}) and (\ref{40})--(\ref{44}), one gets
\beq\label{45}
ds^2=\left(1+\frac{2\Phi}{c^2}\right)c^2dt^2-
\frac{8\delta_{lj}V^l}{c^3}cdtdx^j-
\left(1-\frac{2\Phi}{c^2}\right)\delta_{lj}dx^idx^j\,{.}
\eeq
By calculating the affine connection related to the metric (\ref{45}),
one also obtains the geodesic equations
\beq\label{46}
\ddot{x}^{\alpha}+\Gamma^{\alpha}_{\mu\nu}\dot{x}^{\mu}\dot{x}^{\nu}=0\,
\eeq
where the dots indicate differentiation with respect to the affine parameter. These are the only ingredients necessary to derive the gravitational lens equation for a
beam of light rays  propagating in a weak gravitational field
up to and including terms proportional to $V^l$.

Equations (\ref{46}) yield,
to leading order in $v/c$,

\beq\label{47}
c^2\frac{d^2t}{d\sigma^2}+\frac{2}{c^2}\frac{\partial\Phi}{\partial x^j}c
\frac{dt}{d\sigma}
\frac{dx^j}{d\sigma}-\frac{2}{c^3}\left(\delta_{im}
\frac{\partial V^m}{\partial x^j}+
\delta_{jm}\frac{\partial V^m}{\partial x^i}\right)
\frac{dx^i}{d\sigma}\frac{dx^j}{d\sigma}=0\,{,}
\eeq
$$
\frac{d^2x^k}{d\sigma^2}+
\frac{1}{c^2}\frac{\partial\Phi}{\partial x^j}
\left(c\frac{dt}{d\sigma}\right)^2+
\frac{1}{c^2}\frac{\partial\Phi}{\partial x^k}
\delta{ij}\frac{dx^i}{d\sigma}\frac{dx^j}{d\sigma}-
\frac{2}{c^2}\frac{\partial\Phi}{\partial x^l}\frac{dx^l}{d\sigma}
\frac{dx^k}{d\sigma}+
$$
\beq\label{48}
\frac{4}{c^3}\left(\frac{\partial V^k}{\partial x^j}+
\delta_{jm}\frac{\partial V^m}{\partial x^k}\right)\frac{cdt}{d\sigma}
\frac{dx^i}{d\sigma}=0\,{.}
\eeq
Since for a light ray $ds^2=d\sigma^{2}=0$,
Eq. (\ref{45}) gives,
to order $c^{-3}$,
\beq\label{49}
cdt=\frac{4V^l}{c^3}dx^l+\left(1-\frac{2\Phi}{c^2}\right)dl_{eucl}\,{,}
\eeq
where $dl_{eucl}^2=\delta_{ij}dx^idx^j$ is the Euclidean length interval.
Squaring (\ref{49}) and keeping terms to order $1/c^3$,
one finds
\beq\label{50}
c^2dt^2=\left(1-\frac{4\Phi}{c^2}\right)dl_{eucl}^2+\frac{8V^l}{c^3}dx^ldl_{eucl}\,{.}
\eeq
Inserting (\ref{50}) into (\ref{48}), one gets
\beq\label{51}
\frac{d^2x^k}{d\sigma^2}+
\frac{2}{c^2}\frac{\partial\Phi}{\partial x^k}
\left(\frac{dl_{eucl}}{d\sigma}\right)^2-
\frac{2}{c^2}\frac{\partial\Phi}{\partial x^l}
\frac{dx^l}{d\sigma}\frac{dx^k}{d\sigma}+
\frac{4}{c^3}\left(\frac{\partial V^k}{\partial x^j}-
\delta_{jm}\frac{\partial V^m}{\partial x^k}\right)
\frac{dl_{eucl}}{d\sigma}\frac{dx^j}{d\sigma}=0\,{.}
\eeq
 From (\ref{51}) we derive the equation of the gravitational lens.
Let us consider $l_{eucl}$ as a parameter.
In the approximations used, $d\sigma \sim dl_{eucl}$
and Eq. (\ref{47}) can be written in the form
\beq\label{52}
\frac{d}{d\sigma}\left(\frac{cdt}{d\sigma}+\frac{2\Phi}{c^2}\frac{cdt}{d\sigma}
-\frac{4}{c^3}V^l
\frac{dx^l}{d\sigma}\right)=0\,{,}
\eeq
from which
\beq\label{53}
\frac{cdt}{d\sigma}\left(1+
\frac{2\Phi}{c^2}\right)-\frac{4}{c^3}V^l\frac{dx^l}{d\sigma}
=\mbox{constant}\,{.}
\eeq
The affine parameter can be chosen to make the constant in Eq.(\ref{53})
unity.
Eqs.(\ref{49})
and (\ref{53}) imply that, to lower order in $v/c$,
\beq\label{54}
\frac{dl_{eucl}}{d\sigma}=1\,{.}
\eeq
Therefore, for weak gravitational fields, the vector tangent to
the trajectory of a light ray can be expressed as
\beq\label{55}
\frac{dx^k}{d\sigma}=\frac{dx^k}{dl_{eucl}}\,{,}
\eeq
It is then possible to introduce the vector
\beq\label{56}
e^k=\frac{dx^k}{dl_{eucl}}\,{,}
\eeq
and Eq. (\ref{51}) can be recast in the form
\beq\label{57}
\frac{de^k}{dl_{eucl}}+\frac{2}{c^2}
\frac{\partial \Phi}{\partial x^k}-\frac{2}{c^2}e^ke^l
\frac{\partial \Phi}{\partial x^l}+
\frac{4}{c^3}\left(\frac{\partial V^k}{\partial x^j}-
\frac{\partial V^j}{\partial x^k}\right)e^j=0\,{.}
\eeq
 By using the relation
\beq
\vec{e}\wedge (\nabla\wedge\vec{V})=
\nabla (\vec{e}\cdot \vec{V})-(\vec{e}\cdot\nabla)\vec{V}\,
\eeq
in Eq. (\ref{57}) one finds
\beq\label{58}
\frac{d\vec{e}}{dl_{eucl}}=-
\frac{2}{c^2}[\nabla\Phi-\vec{e}(\vec{e}\cdot\nabla\Phi)]+
\frac{4}{c^3}[\vec{e}\wedge (\nabla\wedge\vec{V})]\,{.}
\eeq
The first term is the component of the gradient of $\Phi$ orthogonal to
the vector $\vec{e}$,
\ie $\nabla_{\perp}\equiv \nabla - \vec{e}(\vec{e}\cdot\nabla)$, while
the second term, is the {\it gravitomagnetic--term}.
Eq.(\ref{58}) reads
\beq\label{59}
\frac{d\vec{e}}{dl_{eucl}}=
-\frac{2}{c^2}\nabla_{\perp}\Phi+\frac{4}{c^3}
[\vec{e}\wedge (\nabla\wedge\vec{V})]\,{,}
\eeq
where $\vec{e}$ is a unit vector.
The deflection angle $\hat{\alpha}$ of a light ray propagating in a weak gravitational field
is given by $\hat{\alpha}=\vec{e}_{in}-\vec{e}_{out}$.
In the general case, one finds from Eq.(\ref{59})
\beq\label{60}
\hat{\alpha}=\frac{2}{c^2}\int \nabla_{\perp}
\Phi dl_{eucl}-\frac{4}{c^3}\int [\vec{e}\wedge (\nabla\wedge\vec{V})]
dl_{eucl}\,{,}
\eeq
where the first term behaves as $\Phi/c^2$ and the second as
$\Phi v/c^3$.

Alternatively,
the equation of a lens can be recovered by using Fermat's principle.
In fact, in classical optics, the light rays follow trajectories
which minimize the optical path $\int n dl$.
One can solve the equation $ds^2=0$ with respect to the temporal coordinate
$dx^0$. The trajectory of a light ray is the extremal of the integral
\beq\label{61}
\int dx^0=\int \left[-\frac{g_{0i}}{g_{00}}\frac{dx^i}{dl_{eucl}}
+\sqrt{\left(\frac{g_{0i}}{g_{00}}\frac{dx^i}{dl_{eucl}}\right)
\left(\frac{g_{0j}}{g_{00}}\frac{dx^j}{dl_{eucl}}\right)
-\frac{g_{ij}}{g_{00}}\frac{dx^i}{dl_{eucl}}\frac{dx^i}{dl_{eucl}}}\right]
dl_{eucl}\,{,}
\eeq
with $dl_{eucl}=\delta_{ij}dx^idx^j$. The integrand in Eq. (\ref{61}) is the refractive index $n$ of a light ray
propagating in a gravitational field. In the weak field approximation,
 one can use Eq. (\ref{49}) to get
\beq\label{63}
\delta\left\{\int \frac{4V^l}{c^3}dx^l+\left(1-\frac{2\Phi}{c^2}\right)
\sqrt{\delta_{ij}dx^idx^j}\right\}=0\,{,}
\eeq
which gives the trajectories of light rays. It therefore follows that
\beq\label{64}
n= \left(1-\frac{2\Phi}{c^2}\right)+\frac{4V^l}{c^3}\frac{dx^l}{dl_{eucl}}\,{.}
\eeq
By using the Euler--Lagrange equations and using
the previous results,
one easly obtains the deflection angle $\alpha$ given by Eq.(\ref{60}).
For small deflection angles and weak gravitational fields, which
are the regimes of practical interest, the true position of a light source
on the sky relative to the position of its image(s) can be defined.
The lens equation can be recast in the form
\beq
\label{lens}
\vec{\theta}-\vec{\theta}_s=
-\left(\frac{D_{ls}}{D_{os}}\right)\vec{{\tilde{\alp}}}(\vec{\theta})=
\hat{\alp}\,,
\eeq
where $\vec{\theta}$ is the position(s) of image(s) with respect to the
optical axis, $\vec{\theta}_s$ the position of the source and
$\vec{{\tilde{\alp}}}$ is the displacement angle. $D_{ls}$ and $D_{os}$
represent respectively the distance between the lens and the source and that
between the observer and the source.

In general, a given image position always corresponds to a
specific source position, whereas a given source position may correspond
to several distinct image positions. For point-mass lenses, the geometry of the system is simplified and we need not
use the full vector equation (\ref{lens}).
In this case, the deflection angle to order $c^{-3}$ is given by
\beq
\label{deflection}
\hat{\alpha}\simeq \frac{4GM}{c^2r_0}-\frac{8GMv}{c^3r_0}\,{,}
\eeq
where $r_0$ is the impact parameter.
 From the definition of red--shift, $\Delta\lambda/\lambda =v/c$,
it follows that the deflection angle can be rewritten as
\beq\label{67}
\hat{\alpha}\simeq \frac{4GM}{c^2r_0}-\frac{8GM}{c^2r_0}
\frac{\Delta\lambda}{\lambda}
=\frac{4GM}{c^2r_0}\left(1-\frac{2\Delta\lambda}{\lambda}\right)\,{.}
\eeq
If $\Delta\lambda/\lambda$ is not negligible, the
microlensing quantities can exhibit an interesting behaviour.

By writing  $r_{0}=\theta D_{ol}$, the lens equation
for a point--mass lens takes the form
\beq
\label{point}
\alp=
\left(\frac{4GM}{c^{2}\theta}\right)\left(\frac{D_{ls}}{D_{os}D_{ol}}\right)
\left(1-\frac{2\Delta\lambda}{\lambda}\right)=
\theta-\theta_{s}\,,
\eeq
which can be rewritten as
\beq
\label{0.5}
\theta^{2}-\theta_{s}\theta-\theta_{E}^{2}=0\,,
\eeq
where
\beq
\label{0.3}
\theta_{E}^{2}=\frac{4GM(\leq r_{E})D_{ls}}{c^{2}D_{ol}D_{os}}
\left(1-\frac{2\Delta\lambda}{\lambda}\right)\,
\eeq
is the square of the Einstein angle defined by $r_{E}=\theta_{E}D_{ol}$.
$\theta_{E}$
depends on the distances involved and the mass of
the deflector. The symbol $M(\leq r_{E})$ signifies that the mass of the lens
must be be contained inside a sphere of radius $r_{E}$.

Before solving the algebraic Eq.(\ref{0.5}), the important parameter {\it magnification} must be discussed. Gravitational lensing preserves
the surface brightness of a source. Then the ratio
of the solid angle $d\Omega_{i}$ covered by the lensed image to that of
the unlensed source $d\Omega_{s}$ gives the flux amplification (magnification)
due to lensing. This is given by the Jacobian of the transformation
matrix between the source and the image(s)
\beq
\label{0.8}
\mu=\frac{d\Omega_{i}}{d\Omega_{s}}=
\left|\mbox{det}\left(\frac{\pa\vec{\theta}_{s}}{\pa\vec{\theta}_{i}}\right)
\right|^{-1}\;.
\eeq
If there is more than a single image, the total magnification
is the sum of all image magnifications.
Considering, as we do, a gravitational point-mass lens
which is
axially symmetric with respect to the line--of--sight, we can use
the scalar angle (\ref{deflection}) and apply Gauss's law
for the total flux. The light deflection reduces to a one--dimensional problem
and Eq.(\ref{0.8}) becomes
\beq
\label{0.9}
\mu=\frac{\theta_{i}d\theta_{i}}{\theta_{s}d\theta_{s}}\,,
\eeq
which can be easily applied \cite{ehlers}.
Let us now solve Eq.(\ref{0.5}). We get
\beq
\label{0.6}
\theta_{\pm}=\frac{\theta_{s}}{2}\pm\sqrt{\frac{\theta_{s}^{2}}{4}+
\theta_{E}^{2}}\;;
\eeq
from which we see that
\beq
\label{0.7}
\theta_{s}=0\;\;\;\;\longrightarrow\;\;\;\;\theta_{\pm}=\pm\theta_{E}\,{.}
\eeq
Because of (\ref{0.3}), the position of the images is therefore shifted by
$\Delta\lambda/\lambda\sim v/c$.
Eqs.(\ref{0.6}), (\ref{0.7}) indicate that we must expect at least
two images from the same source. These lie on the same plane of
the source. It also follows that all quantities where $\theta_{E}$ and $\theta_{\pm}$
appear are modified by $\Delta\lambda/\lambda$.
As discussed above,
it is in general difficult to separate the two images and the  result is a luminosity
enhancement of the source.
The magnification corresponding
to Eq.(\ref{0.6}) is
\beq
\label{0.11}
\mu_{\pm}=\left[1-\left(\frac{\theta_{E}}{\theta_{\pm}}\right)^{4}
\right]^{-1}\,,
\eeq
which shows that when $\theta_{s}$ is zero, the magnification becomes
singular. Physically, this means that when the optical system
source--lens--observer is aligned, we can get a huge magnification.
The total amplification due to both images is
\beq
\label{0.12}
\mu=|\mu_{-}|+|\mu_{+}|=\frac{\chi^{2}+2}{\chi\sqrt{\chi^{2}+4}}\;;
\eeq
where
$\chi=\theta_{s}/\theta_{E}$.
It immediately follows that
${\displaystyle \theta_{s}\leq\theta_{E}\;\;\longrightarrow\;\;\mu\geq
1.34\,,}$
which is the condition for the magnification inside the Einstein ring: a magnification $\mu\sim 1.34$ corresponds to
a magnitude enhancement $\Delta m\sim 0.32$ as required in microlensing
experiments. In other words, when the true position of a light
source lies inside the Einstein ring, the total magnification of the two
images amounts to $\mu\geq 1.34$. This means that the angular
cross
section necessary in order to have significant lensing (\ie $\mu\sim 1.34$ and
$\Delta m \sim 0.32$),  is equal to $\pi\theta_{E}^{2}$ which, from
(\ref{0.3}), is proportional to the mass $M$ of the deflector and to
the ratio of the distances involved.
We can now calculate the optical depth in the case of randomly distributed point--mass lenses. It is possible to estimate the frequency of significant gravitational lensing
events from the observations of distant compact sources. This is equivalent to
considering optical systems with angular sizes
much smaller than $\theta_{E}$. In this situation, the magnification of a
compact source is equal or greater than $1.34$
(since $\theta_{s}<\theta_{E}$) and the probability $P$ of significant
lensing for a randomly located compact source at a distance $D_{os}$
is given by
\beq
\label{0.14}
P=\frac{\pi\theta_{E}^{2}}{4\pi}=\left(\frac{D_{ls}}{D_{os}D_{ol}}\right)
\left(\frac{GM}{c^{2}}\right)\left(1-2\frac{\Delta\lambda}{\lambda}\right)\,,
\eeq
where use has been made of (\ref{0.3}).
Eq. (\ref{0.14}) is linear in
$M$ and therefore holds true also when several
point--mass lenses are
present because the masses can be summed up. Assuming a constant density
for the lens(es) and a static background (this last assumption
surely holds for galactic distances), and averaging over the distances
$D_{ls}, D_{ol}, D_{os}$, the probability (\ref{0.14}) can be interpreted as
the  optical depth $\tau$  for lensing
\cite{ellis},\cite{harwit},\cite{press}.

The considerations given above show that the $c^{-3}$ corrections affect all microlensing quantities,
\ie $\alp, \theta_{E}, \mu, \tau$.
Some estimates are now in order. Let us consider an object receding from us
with a velocity $v\simeq 10^{3}$Km/s. Since$\Delta\lambda/\lambda=v/c$,
the correction is one percent and
could be measured with some difficulty. If the recession velocity
is $v\simeq 10^{4}$Km/s, the correction is ten percent and could in all likelyhood be observed. Far objects like quasars can easily
possess similar recession velocities. It is therefore conceivable that by carrying out observations with precision higher than presently available it would be possible to observe the effects discussed in the case of cosmological microlensing.
Essentially, they should consist in a shift of image positions
and in a reduction of the amplification curve.
Extremely precise observations could, in our opinion, detect
the effects. In a forthcoming paper, we will apply these results to other lens models such as the isothermal sphere or the
disk of galaxies.

\end{document}